\documentclass[twocolumn,showpacs,preprintnumbers,superscriptaddress,prl,amsmath,amssymb,nofootinbib]{revtex4-1}

\usepackage{graphicx}% Include figure files
\usepackage{dcolumn}% Align table columns on decimal point
\usepackage{bm}% bold math
\usepackage{color}
\usepackage{ulem}
\usepackage{gensymb}
\usepackage{braket}
\usepackage{epstopdf}
\usepackage{amsmath}
\usepackage[percent]{overpic}

\begin{document}

\newcommand{\customsection}[1]{\textit{#1.\textemdash}}
\newcommand{\temperaturerange}[3]{#1\,K $\leq$ #2 $\leq$ #3\,K}

\title{Evolution of structural, magnetic and transport properties in MnBi$_{2-x}$Sb$_x$Te$_4$\\}

\author{J.-Q. Yan}
\affiliation{Materials Science and Technology Division, Oak Ridge National Laboratory, Oak Ridge, Tennessee 37831, USA}
\email{yanj@ornl.gov}

\author{S. Okamoto}
\affiliation{Materials Science and Technology Division, Oak Ridge National Laboratory, Oak Ridge, Tennessee 37831, USA}

\author{M. A. McGuire}
\affiliation{Materials Science and Technology Division, Oak Ridge National Laboratory, Oak Ridge, Tennessee 37831, USA}

\author{A. F. May}
\affiliation{Materials Science and Technology Division, Oak Ridge National Laboratory, Oak Ridge, Tennessee 37831, USA}

\author{R. J. McQueeney}
\affiliation{Ames Laboratory and Department of Physics and Astronomy, Iowa State University, Ames, Iowa 50011, USA}

\author{B. C. Sales}
\affiliation{Materials Science and Technology Division, Oak Ridge National Laboratory, Oak Ridge, Tennessee 37831, USA}

\date{\today}

\begin{abstract}
Here we report the evolution of structural, magnetic and transport properties in MnBi$_{2-x}$Sb$_x$Te$_4$ (0$\leq x \leq$2) single crystals. MnSb$_2$Te$_4$, isostructural to MnBi$_2$Te$_4$,  is successfully synthesized in single crystal form. Magnetic measurements suggest an antiferromagnetic order below T$_N$=19\,K for MnSb$_2$Te$_4$ with the magnetic moments aligned along the crystallographic \textit{c}-axis. With increasing Sb content in MnBi$_{2-x}$Sb$_x$Te$_4$, \textit{a}-lattice parameter decreases linearly following Vegard's law while \textit{c}-lattice parameter shows little compositional dependence. The contraction along \textit{a} is caused by the reduction of the Mn-Te-Mn bond angle while the Mn-Te bond length remains nearly constant.  The antiferromagnetic ordering temperature slightly decreases from 24\,K for MnBi$_2$Te$_4$ to 19\,K for MnSb$_2$Te$_4$. More dramatic change was observed for the critical magnetic fields required for the spin-flop transition and the  moment saturation. Both critical fields decrease with increasing Sb content for $x\leq1.72$; a spin-flip transition occurs in MnSb$_2$Te$_4$ at a small field of 3\,kOe applied along \textit{c}-axis. In high magnetic fields, the saturation moment at 2\,K shows significant suppression from 3.56$\mu_B$/Mn for MnBi$_2$Te$_4$ to 1.57$\mu_B$/Mn for MnSb$_2$Te$_4$. Analysis of the magnetization data suggests that both the interlayer magnetic interaction and single ion anisotropy decrease with increasing Sb content for $x\leq1.72$. The partial substitution of Bi by Sb also dramatically affects the transport properties. A crossover from n-type to p-type conducting behavior is observed around $x\approx0.63$.  Our results show close correlation between structural, magnetic and transport properties in MnBi$_{2-x}$Sb$_x$Te$_4$ and that partial substitution of Bi by Sb is an effective approach to fine tuning both the magnetism and transport properties of MnBi$_{2-x}$Sb$_x$Te$_4$.

\end{abstract}

\maketitle

\section{Introduction}

\begin{figure} \centering \includegraphics [width = 0.47\textwidth] {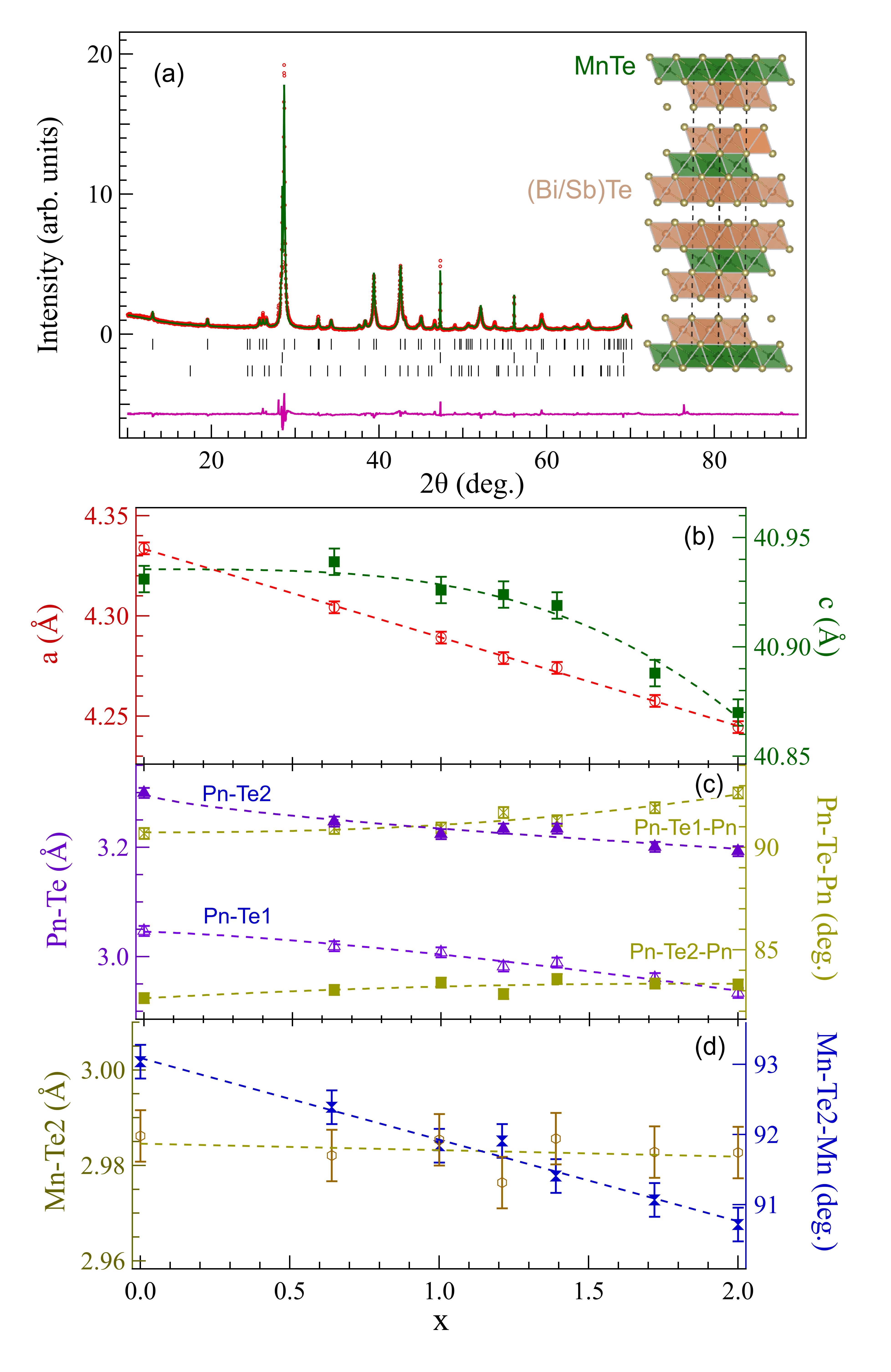}
\caption{(color online) Results of room temperature x-ray powder diffraction. (a) Rietveld fit of powder x-ray diffraction data collected from MnSb$_2$Te$_4$ crystals ground with fine silicon powder. The upper, middle, and lower tick marks locate reflections from MnSb$_2$Te$_4$, silicon, and Sb$_2$Te$_3$, respectively. Inset illustrates the crystal structure for the whole series of MnBi$_{2-x}$Sb$_x$Te$_4$.(b-d) Doping dependence of (b) lattice parameters, (c) Pn-Te bondlength and Pn-Te-Pn bond angle (Pn=pnictogen), and (d) Mn-Te bondlength and Mn-Te-Mn bond angle. The dashed curves are a guide to the eye.}
\label{XRD-1}
\end{figure}

Intrinsic magnetic topological insulators are ideal for realizing exotic quantum states of matter such as an axion insulator and the quantum anomalous Hall effect at elevated temperatures. It has recently been proposed that MnBi$_2$Te$_4$ is the first example of an antiferromagnetic topological insulator,\cite{otrokov2017highly,otrokov2018prediction}
 which has immediately triggered extensive theoretical and experimental studies on bulk,thin films, and thin flakes.\cite{zhang2018topological,chen2019searching,otrokov2019unique,lee2018spin,li2018intrinsic,vidal2019massive,gong2018experimental,cui2019transport,deng2019magnetic}
As shown in the inset of Figure\,\ref{XRD-1}, the rhombohedral crystal structure of MnBi$_2$Te$_4$ can be viewed as inserting one Mn-Te layer into the quintuple-layers of Te-Bi-Te-Bi-Te in Bi$_2$Te$_3$.\cite{lee2013crystal} The antiferromagnetic order below 24\,K of the Mn sublattice has been studied theoretically and experimentally.\cite{eremeev2017competing,otrokov2018prediction,yan2019crystal} Mn$^{2+}$ ions adopt a high-spin $S=5/2$ and order into an A-type antiferromagnetic structure with ferromagnetic layers coupled antiferromagnetically along the \textit{c}-axis. An ordered moment of 4.04(13)$\mu_B$/Mn at 10\,K was found to align along the \textit{c}-axis by neutron diffraction.\cite{yan2019crystal} MnBi$_2$Te$_4$ is thus a unique natural heterostructure of magnetic layers intergrowing with layers of a topological insulator. It is worth mentioning that MnBi$_2$Te$_4$ inherits the van der Waals bonding inbetween the quintuple-layers in Bi$_2$Te$_3$. This feature facilitates the investigation of quantum phenomena in MnBi$_2$Te$_4$ flakes with reduced/controlled thickness employing the exfoliation techniques developed for two-dimensional materials. This is well illustrated by the recent observation of a field-induced quantized anomalous Hall effect\cite{deng2019magnetic} and quantum phase transition from axion insulator to Chern insulator in MnBi$_2$Te$_4$ flakes.\cite{liu2019quantum}

The as-grown MnBi$_2$Te$_4$ single crystals are heavily electron doped possibly due to nonstoichiometry or antisite defects. \cite{otrokov2018prediction,yan2019crystal,zeugner2019chemical,lee2018spin,cui2019transport} A fine tuning of the Fermi level is needed to realize the proposed topological properties. This approach is taken in Cr-doped Bi$_{2-x}$Sb$_x$Te$_3$ where Chang et al  observed the quantum anomalous Hall effect: the randomly distributed Cr provides the ferromagnetism and Sb substitution fine tunes the charge carrier concentration and mobility.\cite{chang2013experimental} Since Mn ions occupy a specific crystallographic site in the layered MnBi$_2$Te$_4$, intuitively we would expect that the Mn-Te layers and Bi-Te layers can be tuned independently with chemical substitutions targeted for different crystallographic sites to optimize the magnetism and electronic band structure separately.  However, the magnetism and transport properties can be intimately coupled and the fine tuning of one property might have a dramatic effect on the other.

In this work, we report a thorough study of the evolution of structural, magnetic, and electrical properties in MnBi$_{2-x}$Sb$_x$Te$_4$. We also report the first successful synthesis and physical properties of MnSb$_2$Te$_4$, which has been proposed\cite{eremeev2017competing} to be isostructural to MnBi$_2$Te$_4$ but has never been experimentally investigated. The existence of MnSb$_2$Te$_4$ makes it possible to synthesize a complete solid solution MnBi$_{2-x}$Sb$_x$Te$_4$. X-ray powder diffraction data reveal the same rhombohedral structure (space group \textit{R-3m}) in the whole composition range. The partial substitution of Bi by Sb is expected to tune the Fermi level without disturbing the magnetic layers. Indeed, we observed dramatic doping effects on the transport properties. A crossover from n-type to p-type conducting behavior is observed around $x\approx0.63$. However, we also observed dramatic doping effects on the structural and magnetic properties suggesting intimate coupling of the structural, magnetic, and electrical properties in MnBi$_{2-x}$Sb$_x$Te$_4$. With increasing Sb content, the lattice parameters, the antiferromagnetic ordering temperature, Weiss constant, saturation moment, the critical fields for spin flop transition and moment saturation all decrease.  Analysis of the magnetization data suggests that both the interlayer magnetic interaction and single ion anisotropy decrease with increasing Sb content for $x\leq1.72$.

\begin{table*}[!ht]
%\begin{table}[htb]
\caption{\label{Table}The structure parameters of  MnBi$_{2-x}$Sb$_x$Te$_4$ obtained from room temperature x-ray powder diffraction patterns.  The space group is $R-3m$ (no. 166). Atomic coordinates are: Mn (0,0,0), Bi (0,0,z), Te1 (0,0,z), and Te2 (0,0,z).}
\begin{tabular}{c|c|c|c|c|c}
\hline
x& $0$& $0.63$ & $1.0$ &$1.39$  & $2$  \\
\hline
a ($\AA$)	&4.3338(4)&4.3043(4)&4.2892(5)&	4.2741(4)	&4.2445(3)\\
c ($\AA$)	&40.931(6)&40.939(5)&40.926(6)&	40.918(5)	&40.870(5)\\
z$_-$Bi(Sb)	&0.4250(6)&0.4246(6)&0.4245(6)&0.4245(6)&0.4250(6)\\
z$_-$Te1	&0.1342(4)&0.1332(4)&0.1329(4)&0.1323(4)&0.1312(4)\\
z$_-$Te2	&0.2935(8)&0.2931(8)&0.2926(8)&	0.2923(8)	&0.2917(8)\\
$\chi^2$	&6.58&3.63&3.35&	2.41	&3.86\\

\hline
\end{tabular}

\end{table*}

\section{Experimental details}

MnBi$_{2-x}$Sb$_x$Te$_4$ single crystals were grown out of a Bi(Sb)-Te flux.\cite{yan2019crystal} The stoichiometry of each composition was determined by elemental analysis on a cleaved surface using a Hitachi TM-3000 tabletop electron microscope equipped with a Bruker Quantax 70 energy dispersive x-ray system. At least two pieces of crystals from each batch were checked. We did not observe any compositional variation for crystals from the same batch or for the same piece of crystal but at different positions. The experimentally determined composition is used in the manuscript. All crystals are plate-like with a typical in-plane dimension of 3mm$\times$4mm and a thickness in the range of 0.1mm-1mm depending on the growth time. The single crystals are soft and can be easily exfoliated, which makes it difficult to collect high quality single crystal x-ray diffraction data. We thus gently ground the single crystals together with a fine powder of silicon and then performed x-ray powder diffraction at room temperature using a PANalytical X’Pert Pro diffractometer with Cu-K$_{\alpha1}$ radiation. The fine silicon powder helps to grind the soft crystals without inducing severe strain.\cite{mcguire2019chemical}

Magnetic properties were measured with a Quantum Design (QD) Magnetic Property Measurement System in the temperature range 2.0\,K$\leq$T$\leq$\,300\,K and in applied magnetic fields up to 70\,kOe. Magnetization in magnetic fields up to 120\,kOe was measured using the ACMS option of a 14\,T QD Physical Property Measurement System (PPMS). The temperature and field dependent electrical resistivity data were collected using a 9\,T QD PPMS.

To gain insights into the electronic structure of MnBi$_{2-x}$Sb$_x$Te$_4$, we carried out density functional theory (DFT) calculations. We used the Vienna ab initio simulation package (VASP)\cite{Kresse1996} with the projector augmented wave method\cite{Kresse1999} and the generalized gradient approximation in the parametrization of Perdew, Burke, and Enzerhof\cite{perdew1996generalized} for exchange-correlation. We do not include $+U$ corrections because these compounds are itinerant semiconductors with strong hybridization between Mn $3d$ states and Te $5p$ states. For Te and Sb, we use the standard potential in the VASP distribution. For Bi (Mn), we use the potential in which the low lying $d$ (semi core $ p$) states are treated as valence states, Bi$_d$ (Mn$_{pv}$).
To accommodate the AFM ordering, we double the experimental structural unit cell along the $c$-axis and then
use a $12\times12\times2$ {\bf k}-point grid with an energy cutoff of 500~eV. The relativistic spin-orbit coupling is turned on, which allows us to analyze the single-ion spin anisotropy in addition to the magnetic coupling.
To avoid the complexity arising from the random distribution of Bi and Sb at $0<x<2$, we only consider two end compounds MnBi$_2$Te$_4$ ($x=0$) and MnSb$_2$Te$_4$ ($x=2$).

\section{Results}

\subsection{Structural properties}

X-ray diffraction from cleaved surfaces of all compositions show the same series of sharp \textit{(00l)} reflections, which suggests all compositions investigated have the same structure. This is further confirmed by the x-ray powder diffraction measurements performed on crystals gently ground with fine silicon powder. Our x-ray diffraction measurements do not suggest any ordering of Bi and Sb in the doped compositions or the formation of other Mn\textit{B}$_{2n}$Te$_{3n+1}$ (\textit{B}=Bi or Sb or mixture) phases with n$>$1.\cite{aliev2019novel} Reitveld refinement suggests 3-5\%wt of Bi$_{2-x}$Sb$_x$Te$_3$ in the ground crystal, which likely comes from the residual flux on the crystal surface. As reported before,\cite{yan2019crystal} ferromagnetic Mn-doped Bi$_{2-x}$Sb$_x$Te$_3$ might affect the magnetic susceptibility at low fields.

Figure\,\ref{XRD-1} shows the diffraction pattern of MnSb$_2$Te$_4$ and a Reitveld refinement as an example.  The results of the Reitveld fit of some selected compositions are summarized in Table I. With increasing Sb content, the \textit{a}-lattice parameter decreases following Vegard's law. The \textit{c}-lattice parameter of MnSb$_2$Te$_4$ is slightly smaller than that of MnBi$_2$Te$_4$.
While the Mn-Te bondlength shows little change with doping, the Mn-Te-Mn bond angle decreases linearly from 93 degree for MnBi$_2$Te$_4$ to 90.75 degree for MnSb$_2$Te$_4$. Both Pn-Te1 (Pn=pnictogen) and Pn-Te2 bonds contract with increasing Sb content. Pn-Te2-Pn bond angle shows little compositional dependence. In contrast, Pn-Te1-Pn bond angle increases gradually with doping from 90.8 degree for MnBi$_2$Te$_4$ to 92.6 degree for MnSb$_2$Te$_4$. The doping induced change in nearest neighbor Mn-Mn distance, i.e., the \textit{a}-lattice parameter, and Mn-Te-Mn bond angle might be of particular importance in understanding the in-plane magnetic interactions.

\subsection{Magnetic properties}

\begin{figure} \centering \includegraphics [width = 0.50\textwidth] {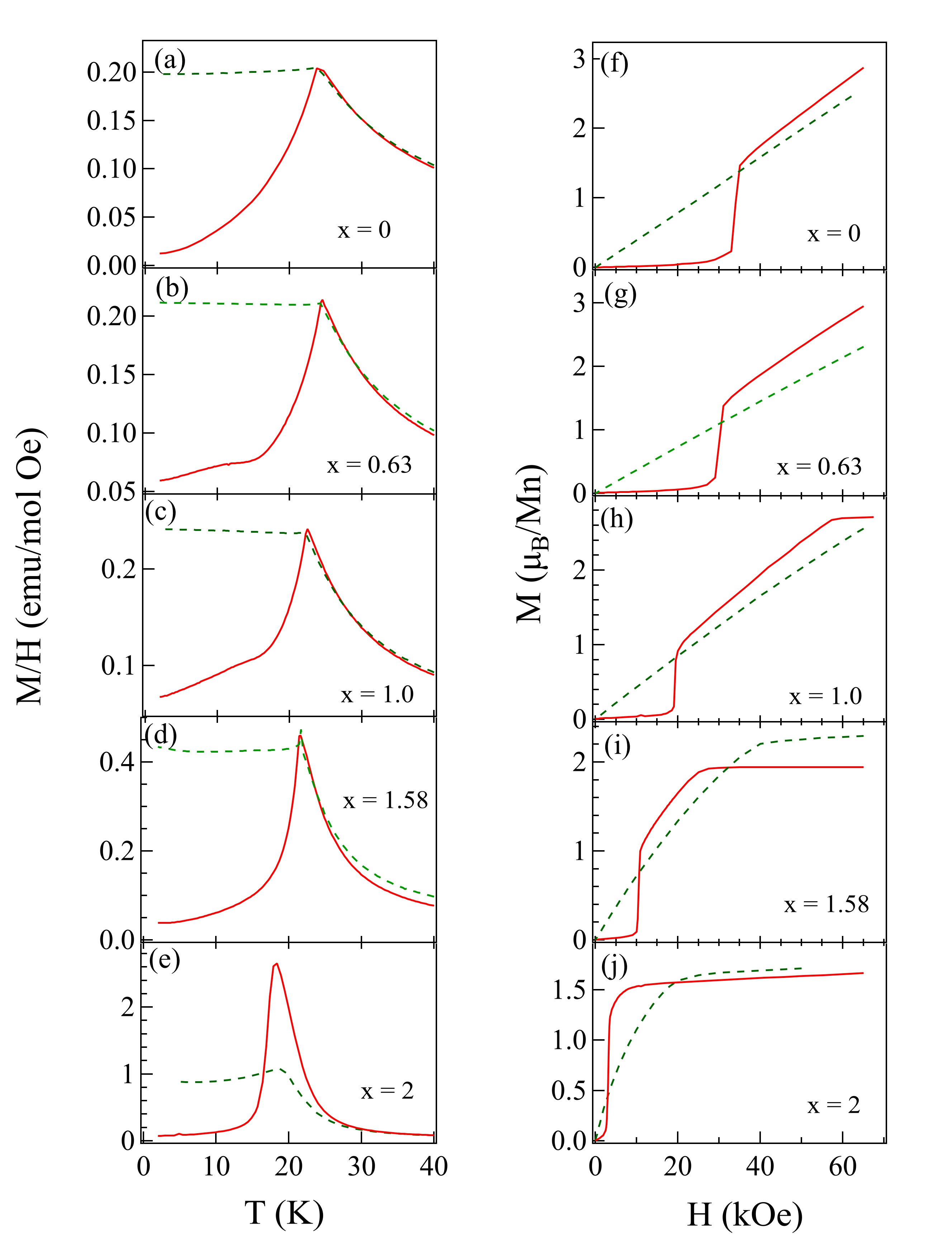}
\caption{(color online) (a-e) Temperature dependence of magnetic susceptibility, M/H, of selected compositions measured in a field of 1\,kOe applied parallel (H//c, solid curve) and perpendicular (H//ab, dash curve) to the crystallographic \textit{c}-axis, respectively. The weak feature around 12\,K in the solid curves in (b) and (c) comes from the ferromagnetic impurity\cite{yan2019crystal}. (f-j) Field dependence of magnetization of selected compositions measured at 2\,K in fields up to 70\,kOe applied parallel (H//c, solid curve) and perpendicular (H//ab, dash curve) to the crystallographic \textit{c}-axis, respectively. Note demagnetization correction is not performed for data with H//c because a demagnetization factor n=1 has negligible effect on the MH curves.}
\label{Mag-1}
\end{figure}

Fifteen compositions of MnBi$_{2-x}$Sb$_x$Te$_4$ with different Bi/Sb ratios were studied. Figure\,\ref{Mag-1}(a-e) show the temperature dependence of the magnetic susceptibility below 40\,K of some selected compositions. The data were collected in a magnetic field of 1\,kOe applied perpendicular and parallel to the crystallographic \textit{c}-axis.  All measurements were done in a field-cooling mode. The data for x=0 are replotted from our previous report.\cite{yan2019crystal}

The magnetic susceptibility is anisotropic below about 30\,K. For MnSb$_2$Te$_4$, a cusp in the M/H curve was observed around 19\,K when the magnetic field is applied along the crystallographic \textit{c}-axis, suggesting the occurrence of a long range magnetic order. The anisotropic behavior of M/H below 19\,K suggests an antiferromagnetic order with moments along the \textit{c}-axis as that found in MnBi$_2$Te$_4$. With increasing Sb content in MnBi$_{2-x}$Sb$_x$Te$_4$, the antiferromagnetic ordering temperature, T$_N$, gradually decreases from 24\,K for MnBi$_2$Te$_4$ to 19\,K for MnSb$_2$Te$_4$. It is worth mentioning that the magnitude of M/H below T$_N$ increases with Sb concentration. This is consistent with the fact that the fields required to flop the spins and saturate the moments decrease with increasing Sb doping as described below. Despite the large difference of M/H below $\approx$30\,K, all compositions have a magnetic susceptibility of around 1.0$\times$10$^{-3}$\,emu/mol.Oe at room temperature.

Figure\,\ref{Mag-1}(f-j) show the field dependence of magnetization, M(H), at 2\,K in magnetic fields up to 70\,kOe.  The isothermal magnetization curves at 2\,K for x=0 and 0.63 were also measured in fields up to 120\,kOe using a QD PPMS (see Figure\,\ref{Hall-1}). For all compositions with x$\leq$1.72, a spin flop transition can be well resolved in M(H) curves when the magnetic field is applied along the \textit{c}-axis. The magnetic field required for the spin-flop transition decreases with increasing Sb content, from 33\,kOe for MnBi$_2$Te$_4$ to 10\,kOe for MnBi$_{0.28}$Sb$_{1.72}$Te$_4$. Meanwhile, the magnetic field required to saturate the moments also decreases with increasing Sb content, from 78\,kOe for MnBi$_2$Te$_4$ to 22\,kOe for MnBi$_{0.28}$Sb$_{1.72}$Te$_4$ for H//c. When the magnetic field is applied perpendicular to the \textit{c}-axis, the magnetization saturates at higher fields and the field required to saturate the moments also decreases with increasing Sb content from 103\,kOe for MnBi$_2$Te$_4$ to 37\,kOe for MnBi$_{0.28}$Sb$_{1.72}$Te$_4$. For MnSb$_2$Te$_4$, only a spin flip transition occurs at 3\,kOe for H//c. For H//ab, the magnetization saturates at 20\,kOe. We notice that the saturation moment drops dramatically with increasing Sb doping from  3.56$\mu_B$/Mn for MnBi$_2$Te$_4$ to 1.57$\mu_B$/Mn for MnSb$_2$Te$_4$.

\begin{figure} \centering \includegraphics [width = 0.47\textwidth] {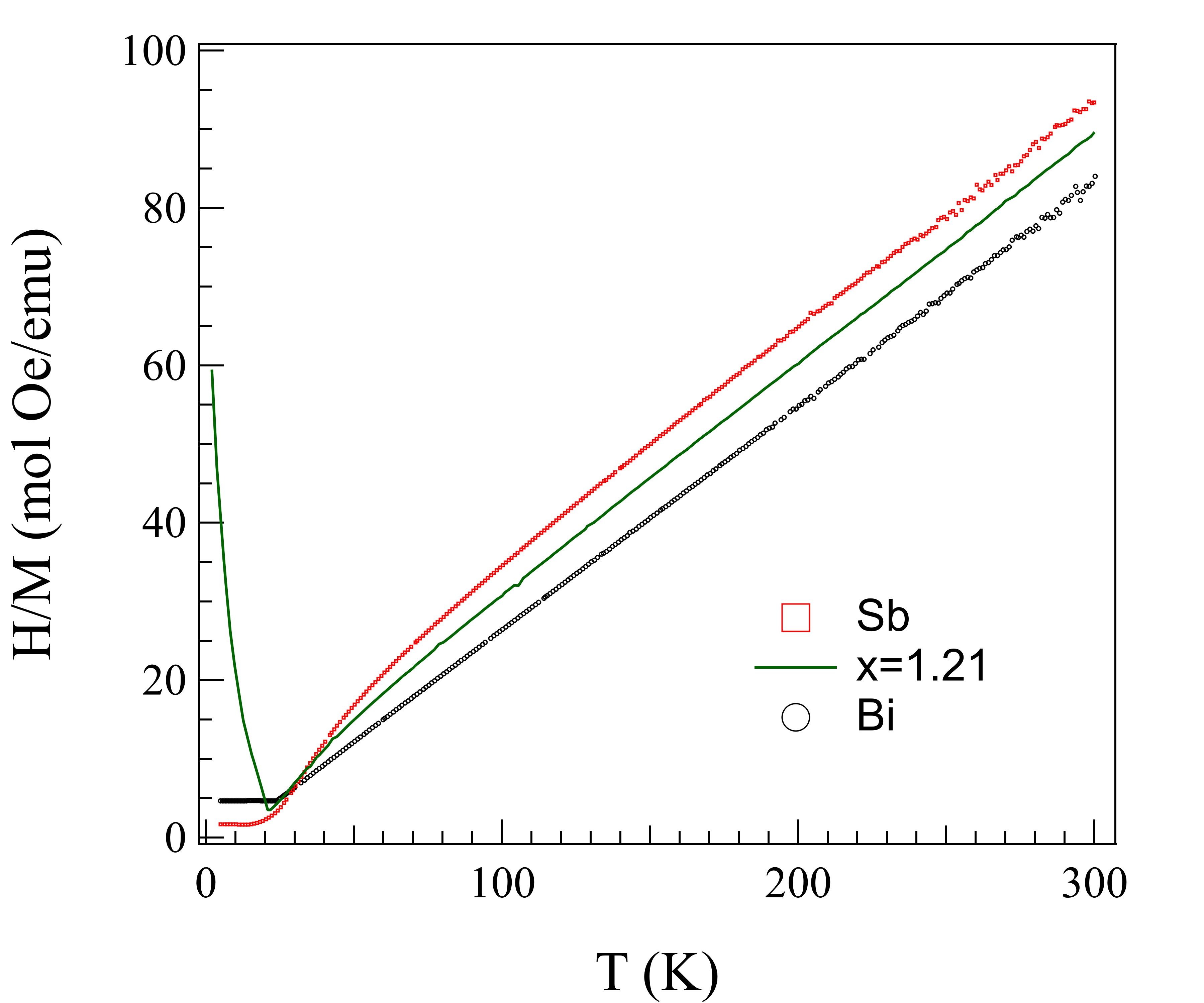}
\caption{(color online) Temperature dependence of reciprocal magnetic susceptibility, H/M, of selected compositions measured in a field of 10\,kOe applied perpendicular (H//ab) to the crystallographic \textit{c}-axis.}
\label{HOverM-1}
\end{figure}

Figure\,\ref{HOverM-1} shows the inverse magnetic susceptibility, H/M, for three compositions. Only the data collected with H//ab were shown because no anisotropic behavior was observed above 30\,K. H/M curves show a linear temperature dependence above 80\,K for all compositions. For MnBi$_2$Te$_4$, the linear temperature dependence extends to T$_N$=24\,K. However, with the partial substitution of Bi by Sb, H/M curves near T$_N$ deviate from the
high temperature linear temperature dependence. As shown in Figure\,\ref{HOverM-1}, the deviation occurs around 80\,K for MnSb$_2$Te$_4$.  This suggests that Sb doping induces short range magnetic correlations above T$_N$ and up to 80\,K for MnSb$_2$Te$_4$. The short range magnetic correlaitons in the paramagnetic state affect the transport properties in this temperature range as presented later. The Curie-Weiss fitting of the data in between 100\,K and 300\,K shows the effective moment is around 5.30\,$\mu_B$/Mn for all compositions without noticeable compositional dependence. However, the Weiss constant gradually changes from 6\,K for MnBi$_2$Te$_4$ to -19\,K for MnSb$_2$Te$_4$. This suggests that the dominant magnetic interaction is changed from ferromagnetic to antiferromagnetic with increasing Sb content, indicating complex competing interactions in MnBi$_{2-x}$Sb$_x$Te$_4$ which deserves detailed studies using techniques such as inelastic neutron scattering. The doping dependence of Weiss constant is summarized in Figure\,\ref{Sum-1} presented later.

The Sb doping induces significant changes in the magnetic properties as well as interlayer coupling and single ion anisotropy as discussed later. The anisotropic magnetic properties suggest an antiferromagnetic order with moments along the \textit{c}-axis in MnBi$_{2-x}$Sb$_x$Te$_4$. All compositions might maintain the same A-type antiferromagnetic order as in MnBi$_2$Te$_4$, yet this needs further confirmation from other measurements such as neutron diffraction. The greatly suppressed saturation moment in MnSb$_2$Te$_4$ might signal strong magnetic fluctuations below T$_N$ or complex spin arrangement in \textit{ab}-plane in the presence of high magnetic fields.

\subsection{Transport properties}

\begin{figure} \centering \includegraphics [width = 0.47\textwidth] {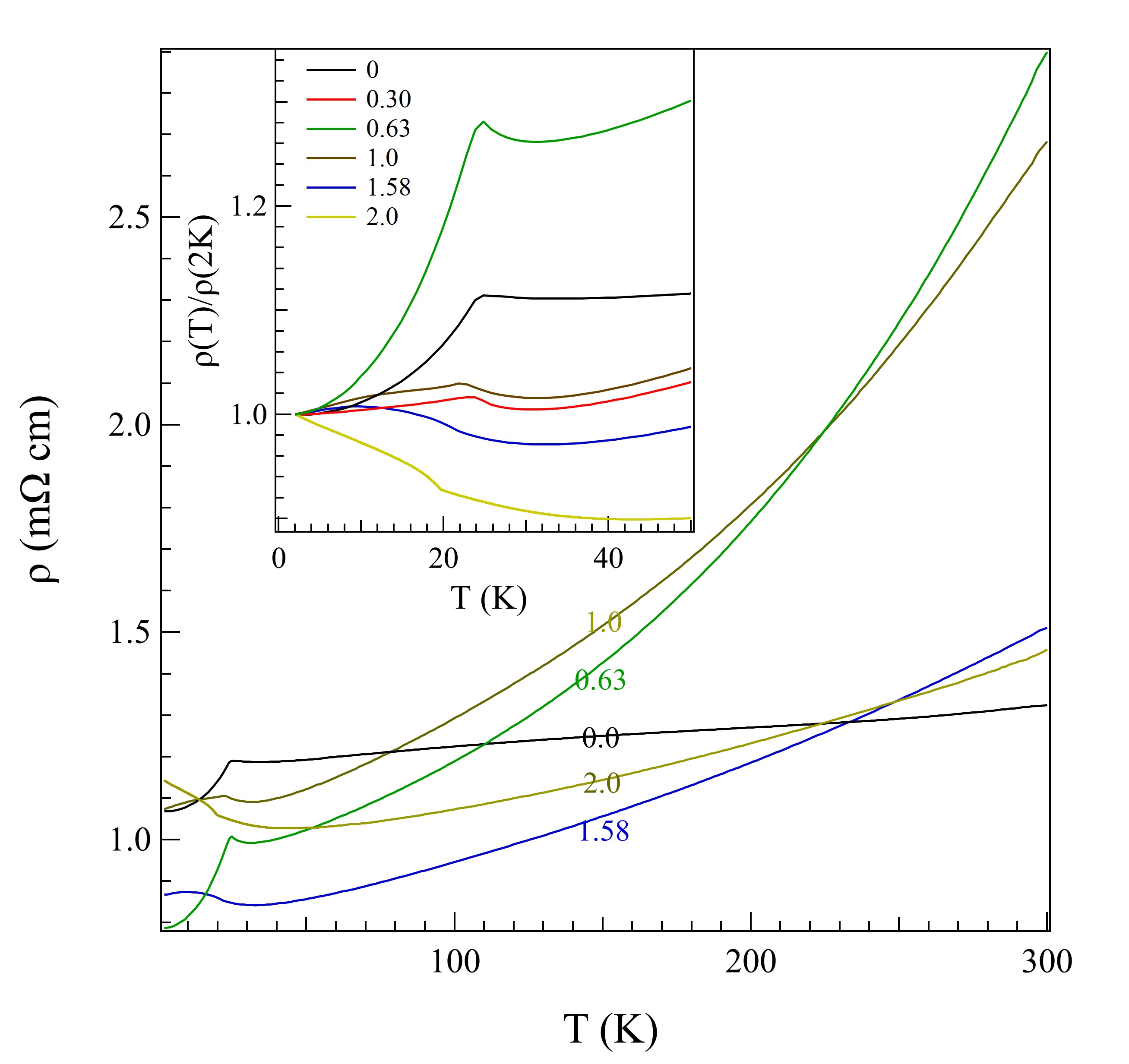}
\caption{(color online) The temperature dependence of in-plane electrical resistivity. Inset shows the normalized in-plane electrical resistivity, $\rho$(T)/$\rho$(2K) to highlight the features at low temperatures.}
\label{RTLT-1}
\end{figure}

The temperature and field dependence of in-plane electrical resistivity was measured in the temperature range 2\,K$\leq$T$\leq$300\,K and in magnetic fields up to 90\,kOe with the electrical current in the \textit{ab}-plane and the field along the \textit{c}-axis.  Figure\,\ref{RTLT-1} shows the temperature dependence of in-plane electrical resistivity, $\rho$(T), of some selected compositions. Above about 50\,K, the $\rho$(T) curves for all compositions show a positive temperature dependence. While $\rho$(300\,K) of most compositions is around 1.2\,m$\Omega$.cm, $\rho$(300\,K) has a value of $\approx$3.0\,m$\Omega$.cm around $x\approx0.63$ where the crossover from n-type to p-type conducting behavior occurs. The compositional dependent transport properties require an understanding of changes in carrier type/concentration, chemical disorder effects, and coupling to the structural and magnetism.

As highlighted in the inset of Figure\,\ref{RTLT-1}, $\rho$(T) of MnBi$_2$Te$_4$ and MnSb$_2$Te$_4$ shows quite different temperature dependence cooling across T$_N$. $\rho$(T) of MnBi$_2$Te$_4$ shows a cusp centered at T$_N$ and is suppressed on cooling below T$_N$ signaling reduced magnetic scattering of electron transport in the magnetically ordered state. In contrast, $\rho$(T) of MnSb$_2$Te$_4$ shows a minimum around 50K, a semiconducting-like temperature dependence below about 50\,K, and a slope change at T$_N$. This temperature dependence suggests the effect of magnetic order on the electrical resistivity in MnSb$_2$Te$_4$ is quite different from that in MnBi$_2$Te$_4$. As presented earlier in Figure\,\ref{HOverM-1}, below about 80\,K, the H/M of MnSb$_2$Te$_4$ deviates from the linear temperature dependence. The semiconducting-like temperature dependence of $\rho$(T) below 50\,K seems to result from the short range magnetic correlations above T$_N$ in MnSb$_2$Te$_4$.  $\rho$(T) of Sb-rich compositions shows a similar temperature dependence as MnSb$_2$Te$_4$ but with a more gradual change cooling across T$_N$ possibly due to the chemical disorder effect. $\rho$(T) of Bi-rich compositions resembles that of MnBi$_2$Te$_4$. As discussed later, our DFT calculations suggest that the doped holes have significant Mn 3d character which drives the  p-type Sb-rich compositions to become more resistive upon cooling across T$_N$.

\begin{figure} \centering \includegraphics [width = 0.47\textwidth] {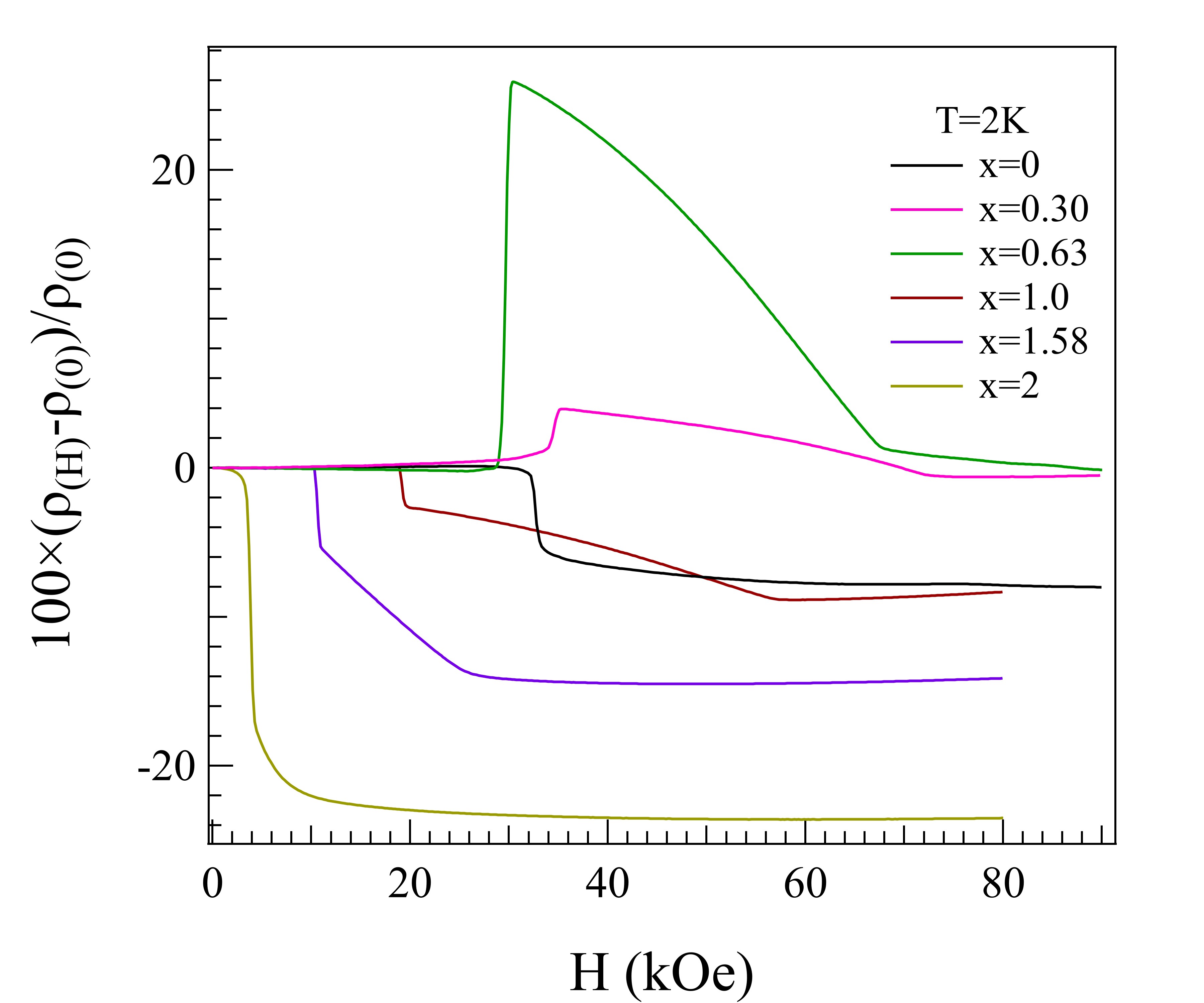}
\caption{(color online) Magnetoresistance of MnBi$_{2-x}$Sb$_x$Te$_4$  at 2\,K in fields up to 90\,kOe. The measurements were performed with the electrical current in \textit{ab}-plane and magnetic fields along the crystallographic \textit{c}-axis. $\rho$(H) was obtained by averaging the data collected in positive and negative fields. Data for MnBi$_2$Te$_4$  was replotted from Ref[\citenum{yan2019crystal}].}
\label{MR2K-1}
\end{figure}

Figure\,\ref{MR2K-1} shows the magnetoresistance (MR) at 2\,K in fields up to 90\,kOe. The measurements were performed with the electrical current in the \textit{ab}-plane and magnetic fields along the crystallographic \textit{c}-axis. $\rho$(H) was obtained by averaging the data collected in positive and negative fields. MnBi$_2$Te$_4$ shows a negative MR with a step-like change at  H$_{c1}\approx$33\,kOe where the spin flop transition occurs and a slope change around H$_{c2}\approx$78\,kOe where the magnetic moment saturates. For all doped compositions investigated in this work,  the step-like change at H$_{c1}$ and the slope change H$_{c2}$ are observed in the MR curves. Both critical fields become smaller with increasing Sb doping. The critical fields determined from MR curves agree with those determined from magnetic measurements. MnSb$_2$Te$_4$ shows a negative MR$\approx$23\% above 3\,kOe. and above this critical field MR shows little field dependence. It is interesting to note that two compositions with x=0.30 and x=0.63 show a positive MR, in contrast to the negative MR of other compositions. As presented later, these two compositions are n-type from the Hall measurements.

\begin{figure} \centering \includegraphics [width = 0.50\textwidth] {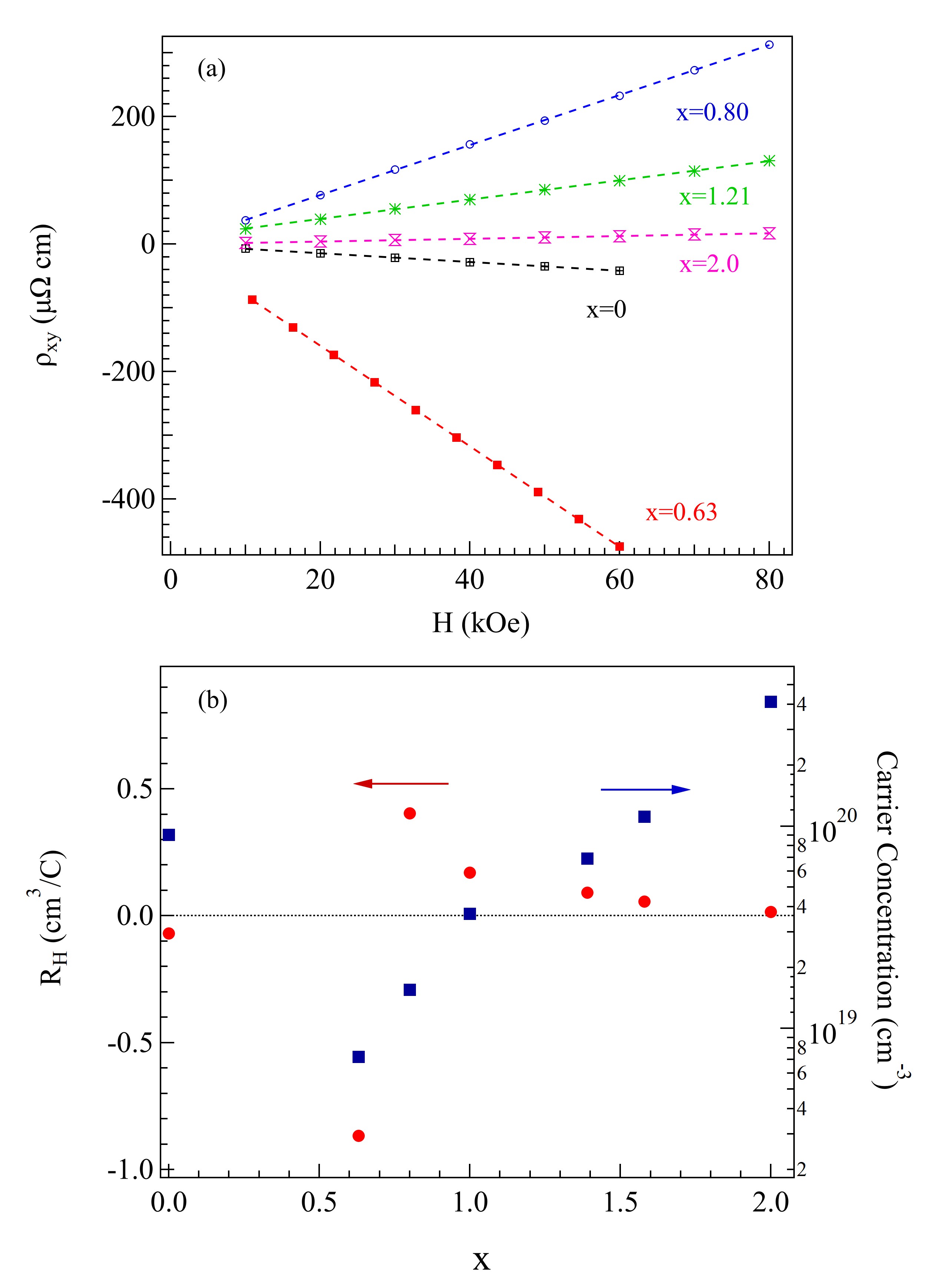}
\caption{(color online) (a) Field dependence of Hall resistivity of MnBi$_{2-x}$Sb$_x$Te$_4$ at room temperature. The measurements were performed with the electrical current in \textit{ab}-plane and magnetic fields along the crystallographic \textit{c}-axis. Dashed lines show the linear fitting. (b) The compositional dependence of Hall coefficient and charge carrier concentration at room temperature. Note that compositions with x$\leq$0.63 are n-type and a crossover from n-type to p-type occurs above x=0.63.}
\label{HallRT-1}
\end{figure}

The Hall resistivity, Hall coefficient, and charge carrier concentration at room temperature are shown in Figure\,\ref{HallRT-1}. The data suggest a transition between n-type and p-type near x=0.63 for our MnBi$_{2-x}$Sb$_x$Te$_4$ crystals.  Near the point of compensation, the Hall coefficient is not particularly meaningful because both types of carriers contribute. However, the analysis of Hall effect data using a single carrier model should be appropriate away from the point of perfect compensation.  For MnBi$_2$Te$_4$, our data suggest a carrier density around 1.3$\times$10$^{20}$ electrons/cm$^3$, while for MnSb$_2$Te$_4$ our data suggest a carrier density of about 5$\times$10$^{20}$ holes/cm$^3$.  A multi-band model may capture some of the physics for compositions near the compensation point.  However, the room temperature Hall effect data measured for these crystals revealed linear or nearly-linear field dependence, thus precluding the use of a multiband model for even the x=0.63 and 0.80 samples.

\begin{figure} \centering \includegraphics [width = 0.50\textwidth] {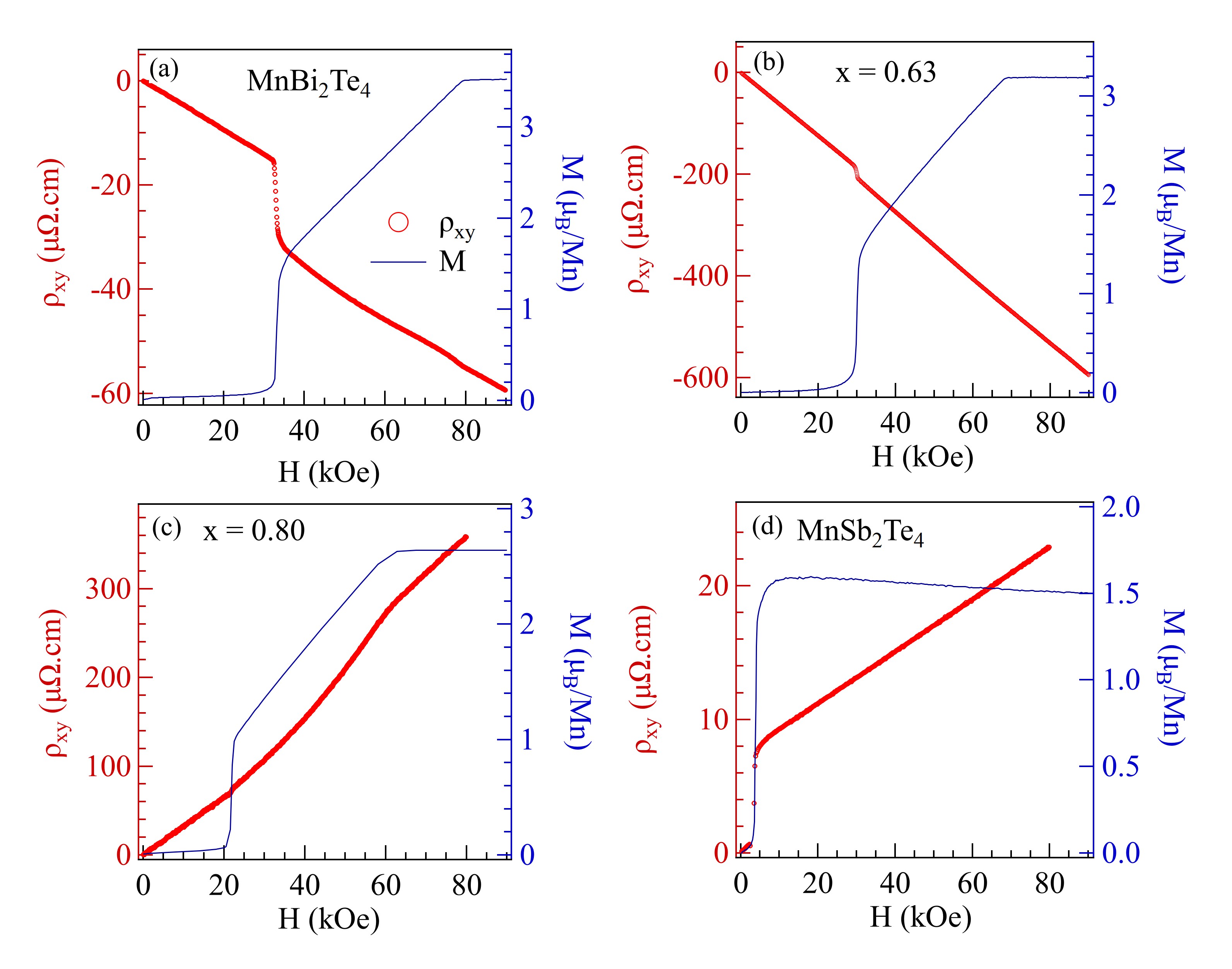}
\caption{(color online) Hall resistivity of MnBi$_{2-x}$Sb$_x$Te$_4$ at 2\,K in fields up to 90\,kOe. (a) x=0, (b) x=0.63. (c) x=0.80. (d) x=2. The measurements were performed with the electrical current in \textit{ab}-plane and magnetic fields along the crystallographic \textit{c}-axis. $\rho_{xy}$ for MnBi$_2$Te$_4$  was replotted from Ref[\citenum{yan2019crystal}]. Magnetization data measured with H//c are also plotted.}
\label{Hall-1}
\end{figure}

We measured the field dependence of Hall resistivity at various temperatures for different compositions. Above 50\,K, a linear field dependence is always observed for those compositions that are away from the compensation point. The Hall coefficient shows a weak temperature dependence.  Upon cooling into the magnetically-ordered state, the magnetic field dependence of the Hall resistivity $\rho_{xy}$ changes dramatically and becomes non-linear.  In particular, an anomalous contribution arises after the sharp increase in M associated with the metamagnetic transition.
Figure\,\ref{Hall-1} shows the Hall resistivity, $\rho_{xy}$, of MnBi$_{2-x}$Sb$_x$Te$_4$ at 2\,K in fields up to 90\,kOe. The measurements were performed with the electrical current in \textit{ab}-plane and magnetic fields along the crystallographic \textit{c}-axis. The magnetization data for each composition collected at 2\,K and in magnetic fields up to 90\,kOe along the \textit{c}-axis were also plotted. In Figure\,\ref{Hall-1} (a-c) where a spin flop transition at H$_{c1}$ is followed by a moment saturation at H$_{c2}$, a step-like change was observed in $\rho_{xy}$ at  H$_{c1}$ and a slope change at  H$_{c2}$. For MnSb$_2$Te$_4$ (see Fig.\,\ref{Hall-1}(d)), a step-like change is observed around 3\,kOe in both $\rho_{xy}$ and M(H) curves.

Below the critical field H$_{c1}$, $\rho_{xy}$  is linear in H, and the Hall coefficient obtained from such data agree well with those obtained at 300K. When M is rising linearly towards saturation, as in Fig.\,\ref{Hall-1}(a-c), the Hall coefficient is nonlinear due to an increasing anomalous contribution.  When the moment reaches a saturation plateau, $\rho_{xy}$ again becomes linear in H. This can be understood as a saturated anomalous contribution and an ordinary Hall coefficient (linear) contribution.  Our data suggest that the ordinary Hall coefficient is essentially unchanged by the spin reorientation. However, a large anomalous contribution is observed. For example, an anomalous Hall resistivity of -20$\mu\Omega$.cm was observed for MnBi$_2$Te$_4$ and +7.3$\mu\Omega$.cm for MnSb$_2$Te$_4$. The anomalous Hall contribution can be of either positive or negative sign, and the sign is not immediately apparent based on other physical properties.  In MnBi$_{2-x}$Sb$_x$Te$_4$ samples, we find that the anomalous contribution has the same sign as the ordinary contribution - a positive anomalous Hall resistivity is observed for hole-doped samples.  This may be revealing something rather significant and, if related to an intrinsic anomalous Hall conductivity, could be investigated theoretically from the band structure of the spin polarized state.

 \begin{figure}
\begin{center}
\includegraphics[width=0.9\columnwidth, clip]{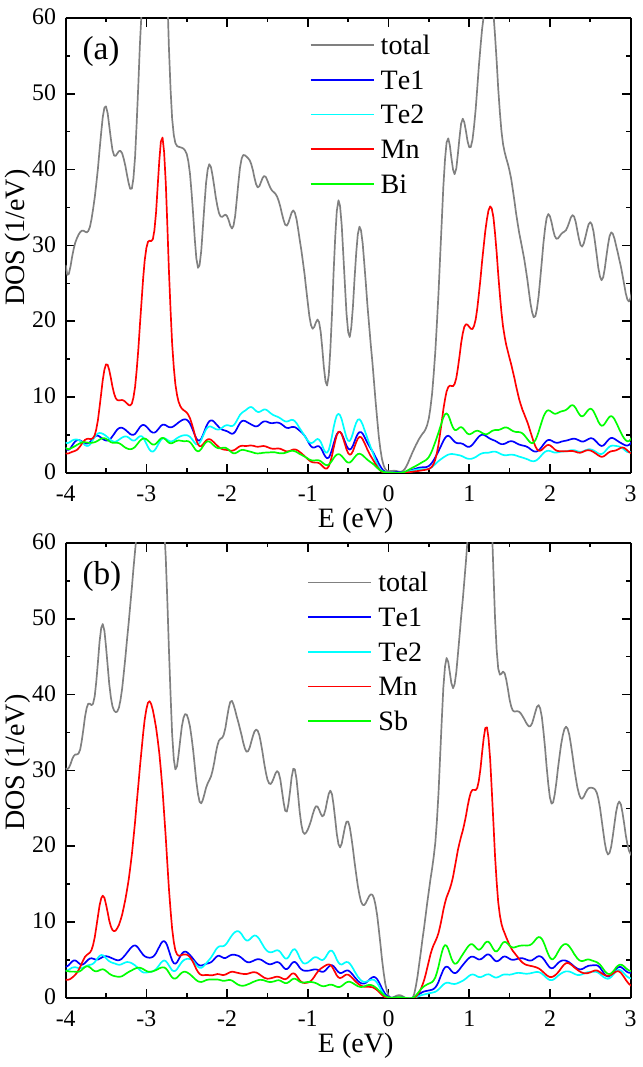}
\caption{Density of states (DOS) of MnBi$_2$Te$_4$ (a) and MnSb$_2$Te$_4$ (b).
Gray lines are the total DOS,
other lines are partial DOS; blue is Te1, light blue Te2, red Mn, and green Bi or Sb.
Density functional theory calculations include antiferromagnetic ordering with the spin-orbit coupling as described in the text.
}
\label{DFT-1}
\end{center}
\end{figure}

\section{DFT calculations}

DFT calculations were performed to understand the experimentally observed difference between MnBi$_2$Te$_4$ and MnSb$_2$Te$_4$. For each compound, we consider layered antiferromagnetic ordering with Mn moments pointing along the $c(a)$ direction, AFM$_{c(a)}$,
and FM ordering with Mn moments pointing along the $c$ direction, FM$_c$.
In both compounds, AFM$_c$ is found to be the ground state. This is consistent with the experimental observations for MnBi$_2$Te$_4$. This can also explain the anisotropic magnetic susceptibility for MnSb$_2$Te$_4$ shown in Figure\,\ref{Mag-1}, although the magnetic structure needs to be determined experimentally.

The ordered M moment is $\sim 4.2 \mu_B$ and the total Mn $d$ charge is $\sim 4.9$ for both compounds. These results are consistent with the valence state of Mn +2. By mapping the total energy of these magnetic ordering to that of a Heisenberg-type model consisting of exchange coupling $J$ between Mn moments and
the single ion anisotropy $D$,
\begin{equation}
H = \sum_{ij} J_{ij} \vec S_i \cdot \vec S_j - D \sum_i |S_i^z|^2
\end{equation}
$J_c S^2\sim 1.21$~meV and $DS^2 \sim 0.74$~meV for MnBi$_2$Te$_4$ and
$J_c S^2\sim 1.18$~meV and $DS^2 \sim 0.45$~meV for MnSb$_2$Te$_4$.
Here, $J_c$ is the nearest-neighbor interlayer coupling, and $J_c>0$ implies it is antiferromagnetic, and $D>0$ implies uniaxial anisotropy. As far as these properties are considered, two compounds look nearly identical.

Figure\,\ref{DFT-1} shows the total and partial density of states of MnBi$_2$Te$_4$ (a) and MnSb$_2$Te$_4$ (b). Our results agree with previous reports.\cite{eremeev2017competing,chen2019searching} Again, two compounds look very similar,
especially Mn $3d$ states appearing at $-3$~eV from majority spins and $1$~eV from minority spins, and
Te $5p$ states dominating at the valence band maximum with some admixture with Mn $3d$ states.
While the low-energy part of the conduction band is dominated by Mn $3d$ minority spin states with some admixture with
 Bi $6p$ or Sb $5p$ states, there is noticeable difference between MnBi$_2$Te$_4$ and MnSb$_2$Te$_4$.
 Bi $6p$ states have the largest weight at the conduction band minimum of MnBi$_2$Te$_4$, while Sb $5p$ states have similar weight with Mn $d$ and Te $5p$ states.
 This difference is caused by heavier Bi than Sb, by which Bi $6p$ states are relatively lower in energy than Sb $5p$ with respect to Mn $3d$ and Te $5p$ states.
While the difference is quite subtle, MnBi$_2$Te$_4$ and MnSb$_2$Te$_4$ are expected to behave differently when doped with electrons;
electric transport may not be so sensitive to AFM ordering in MnBi$_2$Te$_4$ because doped electrons predominantly occupy $6p$ states of Bi,
which is spatially separated from Mn.
On the other hand, when doped with holes, MnBi$_2$Te$_4$ and MnSb$_2$Te$_4$ would behave similarly,
i.e. antiferromagnetic ordering would make electric transport more resistive because doped holes have significant Mn $3d$ character.

Within our DFT, the ordered moment on a Mn site remains unchanged $\sim 4.2 \mu_B$ between MnBi$_2$Te$_4$ and MnSb$_2$Te$_4$.
This is consistent with the experimental value of 4.09$\mu_B$/Mn at 10\,K for MnBi$_2$Te$_4$.\citep{yan2019crystal}
However, the experimental saturation moment shows strong $x$ dependence and is greatly suppressed with increasing Sb doping. This cannot be ascribed to the reduced Mn $3d$ electron density by hole doping. A potential origin of such behavior may be due to the Ruderman–Kittel–Kasuya–Yosida (RKKY) interaction induced by doped holes
\cite{Ruderman1954, Kasuya1956, Yosida1957}.
This could induce magnetic frustrations, resulting in the reduction in the ordered moment, while the effective moment might remain unaffected.

\begin{figure} \centering \includegraphics [width = 0.47\textwidth] {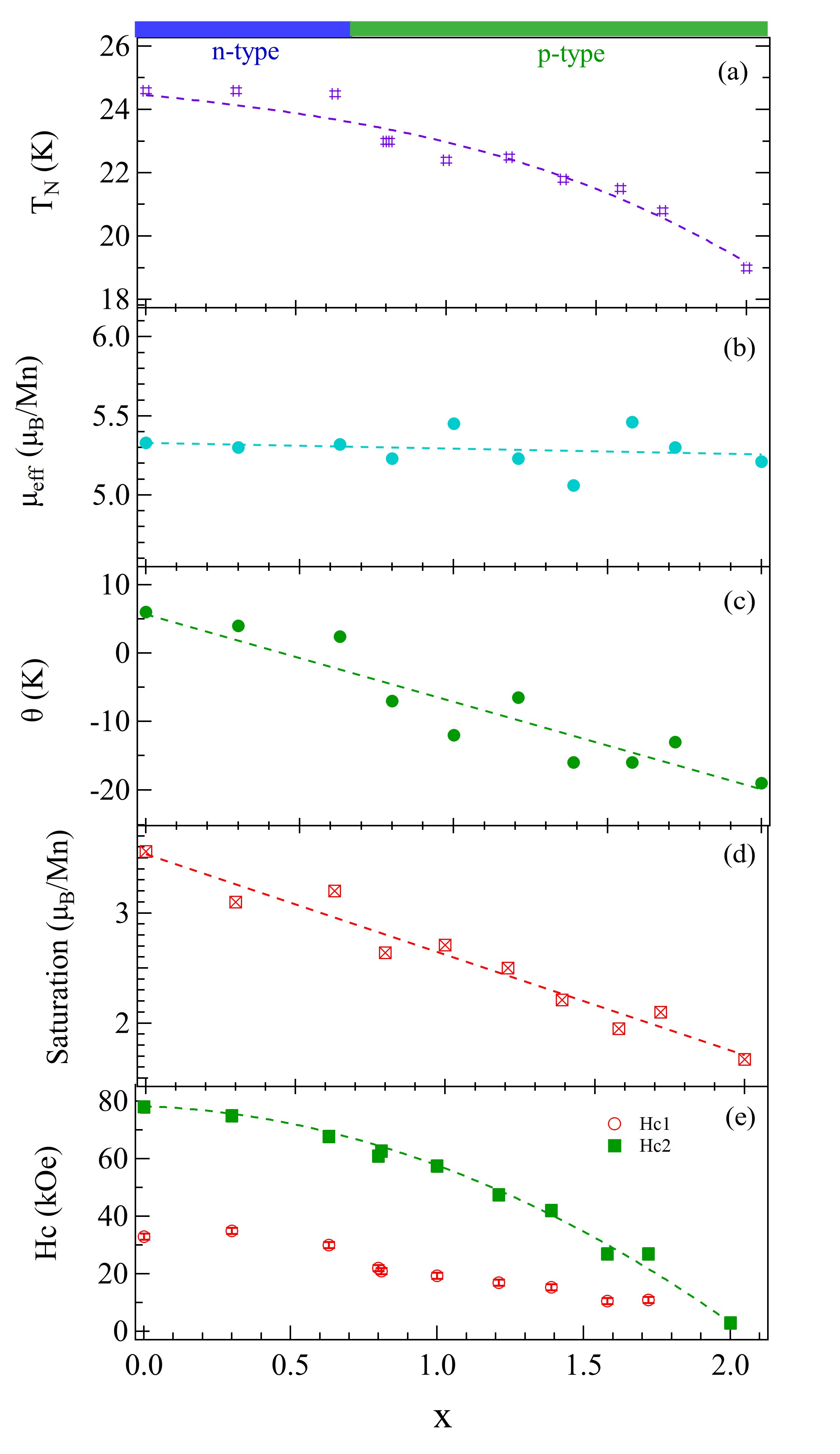}
\caption{(color online) The compositional dependence of (a) Neel temperature, (b) Effective moment obtained from the Curie-Weiss fitting of H/M curves, (c) Weiss constant obtained from the Curie-Weiss fitting of H/M curves, (d) Saturation moment at 2\,K, (e) Critical magnetic fields for spin flop, H$_{c1}$, and moment saturation, H$_{c2}$, determined from magnetic and transport measurements. The dashed curves/lines are a guide for the eye.}
\label{Sum-1}
\end{figure}

\section{Discussion}

Previous DFT calculations show that MnSb$_2$Te$_4$ prefers to crystallize into the same rhombohedral structure as MnBi$_2$Te$_4$.\cite{eremeev2017competing} However, there is no experimental details about the symmetry and atomic parameters of MnSb$_2$Te$_4$ reported before. The successful growth of MnSb$_2$Te$_4$ crystals in this work and our x-ray diffraction studies confirm the DFT predictions and provide the experimental data on the lattice parameters and atomic positions. More importantly, the existence of MnSb$_2$Te$_4$ isostructural to MnBi$_2$Te$_4$ makes possible the synthesis of MnBi$_{2-x}$Sb$_x$Te$_4$  in the whole compositional range. The \textit{a}-lattice constant, thus the nearest neighbour Mn-Mn distance, decreases with increasing Sb content accompanied by the suppression of the Mn-Te-Mn bond angle. Both are expected to modify the magnetic interactions in the hexagonal plane. The direct antiferromagnetic interaction between nearest Mn-Mn neighbours increases with reduced Mn-Mn distance, which competes with the dominant intralayer ferromagnetic interaction and may contribute to the suppressed saturation moment together with the reduced single ion anisotropy.

Figure\,\ref{Sum-1} summarizes the evolution with Sb content of Neel temperature, effective moment, Weiss constant, saturation moment, and the critical fields. The magnetic field required for the flop transition is labeled as H$_{c1}$ and the larger field to saturate the moment H$_{c2}$. The colored bars in Figure\,\ref{Sum-1}(a) were plotted to highlight that the crossover from n-type to p-type conduction occurs around a critical Sb content of x$_c$=0.63. However, no abrupt change or obvious anomaly around x$_c$ was observed for other parameters plotted in Figure\,\ref{Sum-1}. This indicates that around the critical composition, the type and concentration of charge carriers can be finely tuned possibly without inducing any abrupt change of the magnetism.

The effective moments obtained from the Curie-Weiss fitting of H/M curves of different compositions are all around 5.3$\pm$0.1\,$\mu_B$/Mn and show no noticeable composition dependence. This effective moment is as expected for a high spin Mn$^{2+}$ with an electronic configuration of t$_{2g}^3$e$_g^2$. The absence of noticeable compositional dependence of $\mu_{eff}$ also indicates that there is not a valence or spin state change in MnBi$_{2-x}$Sb$_x$Te$_4$.

The Weiss constant for MnBi$_2$Te$_4$ is 6\,K, which is consistent with that reported by Lee et al\cite{lee2018spin}. With increasing Sb content in MnBi$_{2-x}$Sb$_x$Te$_4$, the Weiss constant gradually decreases to -19\,K for MnSb$_2$Te$_4$. As pointed out by Otrokov et al, the temperature independent $\chi_0$ affects the Curie-Weiss fitting as well as the magnitude of Weiss constant. However, reasonable fitting with $\chi_0$ included also gives an negative Weiss constant for MnSb$_2$Te$_4$, which suggests the dominant antiferromagnetic interactions in the paramagnetic state of MnSb$_2$Te$_4$. This is in sharp contrast to the ferromagnetic interactions dominant in MnBi$_2$Te$_4$ and may have profound effect on the physical properties. The ferromagnetic fluctuations in MnBi$_2$Te$_4$ have been proposed to break time-reversal symmetry inducing the gap opening of surface states. It would be interesting to check with ARPES measurements whether such a gap opening of surface states also occurs in MnSb$_2$Te$_4$ below and above T$_N$.

The saturation moments at 2\,K of different compositions are summarized in Figure\,\ref{Sum-1} (d). The saturation moment is drastically suppressed by the replacement of Bi by Sb.  At 2\,K, a saturation moment of 3.56$\mu_B$/Mn was observed for  MnBi$_2$Te$_4$.  Surprisingly, a much smaller saturation moment of 1.67$\mu_B$/Mn was observed for MnSb$_2$Te$_4$. The saturation moment decreases linearly with increasing Sb content. For high spin Mn$^{2+}$ with an electronic configuration of t$^3$e$^2$, a saturation moment of 5$\mu_B$/Mn  is expected.
It is worth mentioning that our x-ray powder diffraction found only a few percent of Sb$_2$Te$_3$ likely from the residual flux on the crystals. This amount of nonmagnetic impurity is not enough to induce such a large reduction of saturation moment. We further measured magnetization at 2\,K in magnetic fields up to 120\,kOe without finding any other field-induced changes. In magnetic fields above H$_{c2}$, the magnetization shows little field dependence up to 120\,kOe. We also measured M/H of MnSb$_2$Te$_4$ in a field of 50\,kOe and M/H tends to saturate below 5\,K. All these suggest the suppression of saturation moment in Sb-bearing compositions is an intrinsic behavior, though the magnetic structure should be verified to exclude more complex spin structures causing the apparent difference in saturation moment..

Our x-ray diffraction (see Fig.\,\ref{XRD-1}) shows that Mn-Te bond length has little doping dependence, which implies little change in the Mn-Te bond covalency. However, the nearest neighbour Mn-Mn distance is reduced with increasing Sb content. It is expected that the enhanced direct antiferromagnetic interactions would compete with the dominant nearest neighbour ferromagnetic interaction. Therefore, the reduced saturation moment can result from the magnetic fluctuations in the magnetically ordered state. If this scenario is right, we would expect the ordered moment of MnSb$_2$Te$_4$ is also smaller than that of MnBi$_2$Te$_4$. Neutron diffraction is needed to investigate the magnetic structure and ordered moment in MnSb$_2$Te$_4$ and maybe the evolution with Sb doping especially if MnSb$_2$Te$_4$ has a different magnetic structure other than the simple A-type antiferromagnetism observed in MnBi$_2$Te$_4$. On the other hand, it would be interesting to investiagate the magnetization in much higher magnetic fields for possible field induced transitions. This is especially true for MnSb$_2$Te$_4$ with such a small saturation moment. 

\begin{figure} \centering \includegraphics [width = 0.47\textwidth] {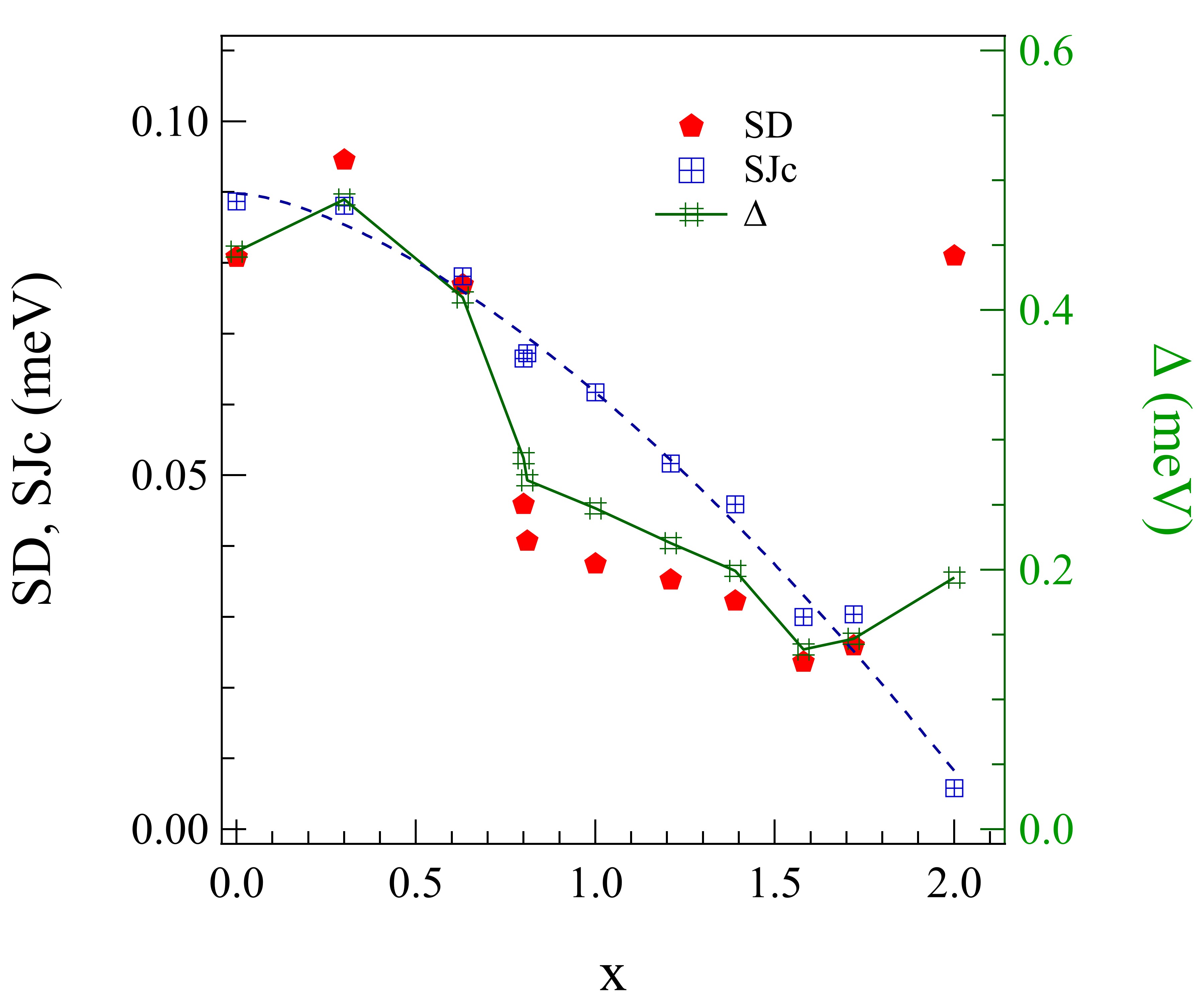}
\caption{(color online) Compositional dependence of single ion anisotropy, SD, and interlayer magnetic interaction, SJc, and spin gap, $\bigtriangleup$. The solid and dashed curves are a guide to the eye.}
\label{SDSJc-1}
\end{figure}

Figure\,\ref{Sum-1} (e) shows the critical magnetic fields for spin flop, H$_{c1}$, and moment saturation, H$_{c2}$, determined from magnetic and transport measurements with magnetic fields applied along the crystallographic \textit{c}-axis. Both critical fields decrease with increasing Sb content. For MnSb$_2$Te$_4$, a magnetic field as small as 3\,kOe is enough to saturate the magnetization. For a uniaxial antiferromagnet, the spin-flop field and saturation field can be written in terms of the interlayer antiferromagnetic exchange (Jc) and single-ion anisotropy (D) as g$\mu_B$H$_{c1}$=2SD$\sqrt{zJc/D-1}$ and g$\mu_B$H$_{c2}$=2SD(zJc/D-1), respectively, where g=2, z=6 is the coordination number for Mn to other Mn in layers above and below. Therefore, Jc and D can be calculated as
SD=(1/2)g$\mu_B$H$_{c1}$(H$_{c1}$/H$_{c2}$), and  SJc=$\frac{1}{2z}$g$\mu_B$H$_{c2}$[(H$_{c1}$/H$_{c2}$)$^2$+1], respectively. The calculated SD and SJc are plotted in Figure\,\ref{SDSJc-1}. For MnSb$_2$Te$_4$ with only a spin-flip transition at a field of H$_{c}$, SJc=g$\mu_B$H$_{c}$/z. SD of MnSb$_2$Te$_4$ can be estimated as SD=(1/2)g$\mu_B$H$_c^*$-zSJc, where H$_c^*$ is the critical field applied perpendicular to the \textit{c}-axis to saturate the magnetic moment .

Both SD and SJc decrease with increasing Sb content for $x\leq1.72$ although they show distinct compositional dependence. It is interesting to note that SJc decreases monotonically with increasing Sb content, although \textit{c}-lattice shows little doping dependence. The step-like change of SD around x=0.70 signals two different compositional dependence of SD: below x=0.63, SD shows little doping dependence; for $0.80\leq x \leq1.72$, SD decreases monotonically with increasing Sb content. This corresponds with the change in carrier type and more compositions will be studied to reveal the details of the doping dependence of SD in the composition range 0.63$<$x$<$0.80.

In the Sb-rich composition around x=1.72, SD and SJc are comparable to each other. For MnSb$_2$Te$_4$, magnetization saturates in a field of 3\,kOe at 2\,K. The critical composition where the spin flop transition disappears is in 1.72$<$x$\leq$2. Above x=1.72, SJc/SD is smaller than 2/z required for the spin flop transition, and therefore, magnetization saturates directly with increasing field. As shown in Fig.\,\ref{SDSJc-1}, SD is one order of magnitude larger than SJc for MnSb$_2$Te$_4$. The magnitude of interlayer coupling relative to the single ion anisotropy determines the different magnetic properties between MnSb$_2$Te$_4$ and MnBi$_2$Te$_4$. We noticed that for MnSb$_2$Te$_4$, the SJc and SD estimated here are different from our DFT calculations. This indicates that Sb doping has  complex effect on the magnetic interactions and/or the magnetic structure of MnSb$_2$Te$_4$ may not be a simple A-type antiferromagnetism.

The spin gap can be calculated as $\bigtriangleup$=2SD$\sqrt{zSJc/SD+1}$. Figure\,\ref{SDSJc-1} also shows the doping dependence of spin gap. The spin gap in MnBi$_2$Te$_4$ is about 0.45\,meV and it decreases with increasing Sb content to about 0.15\,meV for x=1.72. Above x=1.72, the spin gap slightly increases to 0.20\,meV for MnSb$_2$Te$_4$. These need to be confirmed experimentally by, for example, inelastic neutron scattering.

\begin{figure} \centering \includegraphics [width = 0.47\textwidth] {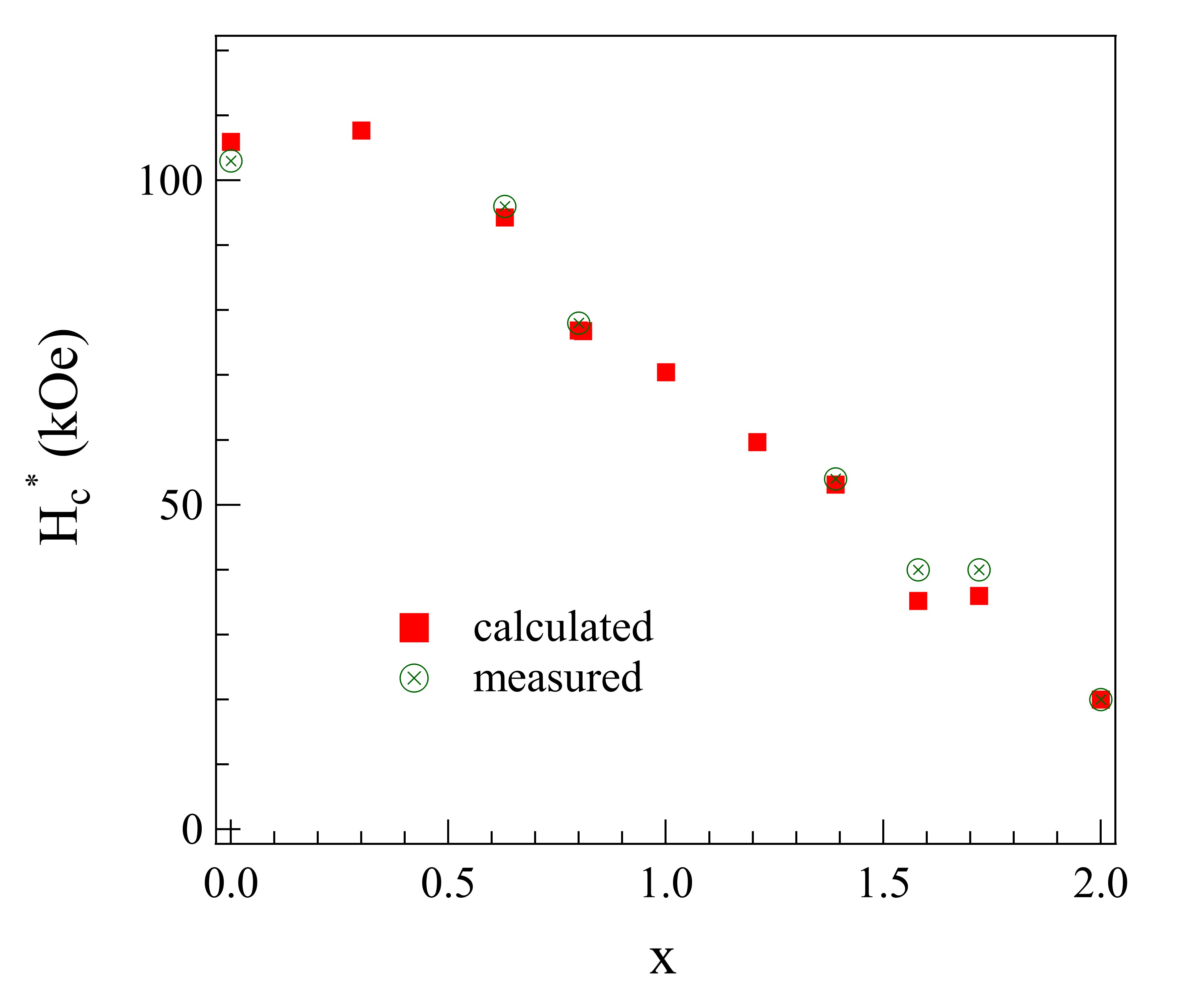}
\caption{(color online) The critical magnetic fields required to saturate magnetization when field is applied perpendicular to the crystallographic \textit{c}-axis. The calculated values are obtained using experimental results to obtain g$\mu_B$H$_c^*$=2(zSJc+SD).}
\label{Hcab-1}
\end{figure}

With the knowledge of SD and SJc, the magnetic field required to saturate magnetization when field is applied perpendicular to the \textit{c}-axis can be calculated as g$\mu_B$H$_c^*$=2(zSJc+SD). As shown in Figure\,\ref{Hcab-1}, the calculated and measured critical fields agree well for all compositions.

\section{Summary}
In summary, we study the evolution of structural, magnetic, and transport properties in the whole compositional range of MnBi$_{2-x}$Sb$_x$Te$_4$. Our results show close correlation between the structural, magnetic, and transport properties in this system. With increasing Sb content, the lattice parameters, the antiferromagnetic ordering temperature, Weiss constant, saturation moment at 2\,K, the critical fields for spin-flop transition and moment saturation decrease. Data analyses suggest that the interlayer exchange coupling, single ion anisotropy, and the magnon gap also decrease with the substitution of Bi by Sb. The significant reduction in the saturation moment at 2\,K cannot be explained simply by the different Mn-Te bond covalency. Instead, the enhanced direct interactions in \textit{ab}-plane due to the reduced Mn-Mn bond length and also the RKKY interactions induced by doped holes should be considered. The modified in-plane magnetic interactions, together with reduced interlayer coupling and single ion anisotropy with Sb doping, might induce strong magnetic fluctuations even in the magnetically ordered state if the same A-type antiferromagnetic strucutre is maintained for the whole system. All these need to be further investigated with advanced techniques such as neutron scattering.

Near x$_c$=0.63, a transition from n-type to p-type conduction is observed in our crystals. Further careful optimization of growth parameters and fine tuning of the chemical compositions are needed to control the magnetism, the amount of lattice defects, charge carrier concentration and mobility to facilitate the realization of theoretically predicted topological properties.

\section{Acknowledgment}

This work was supported by the U.S. Department of Energy, Office of Science, Basic Energy Sciences, Materials Sciences and Engineering Division. Ames Laboratory is operated for the U.S. Department of Energy by Iowa State University under Contract No. DE-AC02-07CH11358.

 This manuscript has been authored by UT-Battelle, LLC, under Contract No.
DE-AC0500OR22725 with the U.S. Department of Energy. The United States
Government retains and the publisher, by accepting the article for publication,
acknowledges that the United States Government retains a non-exclusive, paid-up,
irrevocable, world-wide license to publish or reproduce the published form of this
manuscript, or allow others to do so, for the United States Government purposes.
The Department of Energy will provide public access to these results of federally
sponsored research in accordance with the DOE Public Access Plan (http://energy.gov/
downloads/doe-public-access-plan).

\section{references}
%\bibliographystyle{apsrev4-1}
%\bibliography{bibfile}


\end{document}